\documentclass[11pt]{article}

\RequirePackage[T1]{fontenc}
\usepackage[utf8]{inputenc}

\usepackage[height=8.85in,width=6.45in]{geometry}
\renewcommand{\baselinestretch}{1.2}

\usepackage{titlesec}

\titleformat*{\section}{\Large\bfseries}
\titleformat*{\subsection}{\bfseries}

\usepackage[svgnames]{xcolor}
\definecolor{refcolor}{rgb}{0.3,0.3,1}
\definecolor{blue}{rgb}{0.1, 0.3, 0.8}

\usepackage[
  bookmarks=false,
  colorlinks=true,
  citecolor=refcolor,
  linkcolor=refcolor,
  urlcolor=magenta
]{hyperref}

\usepackage[hang]{caption}
\usepackage{subcaption}

\usepackage[hang,bottom]{footmisc}
\usepackage{appendix}

\usepackage{amsmath}
\numberwithin{equation}{section}

\usepackage{amssymb}
\usepackage{bm} % bold symbols in math mode
\usepackage{braket}
\usepackage{mathdots} % ddots, vdots
\usepackage{mathtools} % coloneqq etc.
\usepackage{nicematrix}
\usepackage{extarrows}

\usepackage{graphicx}
\usepackage{pifont} % additional symbols
\usepackage{marvosym} % additional symbols

\usepackage{ascmac} % screen, itembox etc.

\usepackage{array}
\usepackage{longtable}
\usepackage{arydshln}
\usepackage{multirow}

\usepackage{tikz}
\usetikzlibrary{arrows.meta}

\title{\Large\textbf{%
	Sub-extensive non-stabilizerness\\%
	in the Dyck-Fredkin spin chain\\[20pt]%
}}

\author{\large%
	Yasunori Lee%
}

\date{\small\emph{%
	International Center for Elementary Particle Physics,\\%
	University of Tokyo, Bunkyo, Tokyo 113-0033, Japan\phantom{,}%
}}

\begin{document}

\begin{titlepage}

\maketitle

\vspace{-40pt}

\begin{abstract}
	\normalsize
	The stabilizer Rényi entropy is a quantitative measure of non-stabilizerness, or~\textit{magic},
	and has typically been found to scale extensively with system size $N$ (i.e., $\Theta(N)$) for a variety of many-body quantum states.
	In this note, we study the stabilizer Rényi entropy of the ground state of the spin-$\frac{1}{2}$ Dyck-Fredkin chain and its $t$-deformation,
	a~local frustration-free model with unusual spectral-gap scaling.
	Exploiting the combinatorial structure, we carry out numerically exact finite-size calculations,
	which indicate asymptotic behavior depending on $t$:
	$\Theta(N)$ for $t<1$, 
	$\Theta(\log N)$ at $t=1$, and
	$\Theta(1)$ for $t>1$.
	The scaling at $t=1$ could be another manifestation of the unconventional criticality of the model,
	while the contrast with the behavior of the entanglement entropy suggests that
	non-stabilizerness might provide a new window into quantum many-body systems.
\end{abstract}

\thispagestyle{empty}

\renewcommand*{\thefootnote}{\fnsymbol{footnote}}

\end{titlepage}

\newpage

\renewcommand{\baselinestretch}{0.8}
\tableofcontents

\renewcommand{\baselinestretch}{1.3}
\setcounter{footnote}{0}
\setcounter{page}{1}

%%%%%%%%%%%%%%%%%%%%%%%%%%%%%%%%%%%%%%%%%%%%%
\section{Introduction}
%%%%%%%%%%%%%%%%%%%%%%%%%%%%%%%%%%%%%%%%%%%%%
One natural definition of the ``complexity'' of a quantum state is the difficulty of realizing it.
From the perspective of quantum computation,
there is in fact a special class of operations that, even when used arbitrarily many times, 
can only prepare a proper subset of all quantum states.
This distinguishes so-called stabilizer states from those that are not, 
and further motivates the notion of non-stabilizerness,
which characterizes the extent to which a quantum state differs from stabilizer states.
Among various quantitative measures of non-stabilizerness,
the stabilizer Rényi entropy (SRE) \cite{LeoneOlivieroHamma:2021} 
has recently attracted particular attention due in part to its computational accessibility,
and has begun to be used to study quantum many-body systems \cite{
LiuWinter:2020}.

For many classes of many-body quantum states,
the scaling behavior of the SRE is found to be extensive 
in system size (see, e.g.,
\cite{
OlivieroLeoneHamma:2022,
HaugPiroli:2022,
TarabungaCastelnovo:2023,
TurkeshiDymarskySierant:2023,
FalcãoTarabungaFrauTirritoZakrzewskiDalmonte:2024,
ColluraDeNardisAlbaLami:2024}).
There is even a theoretical account of this behavior in the case of one-dimensional critical states
described by conformal field theory (CFT) \cite{HoshinoOshikawaAshida:2025,HoshinoAshida:2025}.
However, there appears to be no fundamental reason to rule out (non-trivial) sub-extensive scaling of the SRE,
and in fact there \textit{are} some classes of quantum states exhibiting
logarithmic \cite{OdavićHaugTorreHammaFranchiniGiampaolo:2022} or (asymptotically) non-zero constant~\cite{ChenYanZhou:2023} scaling,
while the physical implications of such unusual scaling remain largely unclear.

In this note, we examine the (Dyck-)Fredkin spin chain \cite{DellAnnaSalbergerBarbieroTrombettoniKorepin:2016,SalbergerKorepin:2016},
a local frustration-free model with unconventional spectral properties, and find that
the SRE of the ground state scales logarithmically with system size.
Furthermore, upon deformation of the model \cite{SalbergerUdagawaZhangKatsuraKlichKorepin:2016},
the SRE exhibits constant scaling on one side of the undeformed point and extensive scaling on the other side,
in contrast to the entanglement entropy.
The apparent coincidence between the unusual scaling of the SRE and that of the spectral gap
may hint at hitherto unknown implications of non-stabilizerness,
and the absence of a counterpart in the entanglement entropy
suggests the unique potential of non-stabilizerness as a probe of quantum many-body systems.

%%%%%%%%%%%%%%%%%%%%%%%%%%%%%%%%%%%%%%%%%%%%%
\section{Preliminaries}\label{sec:preliminaries}
%%%%%%%%%%%%%%%%%%%%%%%%%%%%%%%%%%%%%%%%%%%%%
\subsection{Stabilizer Rényi entropy}\label{subsec:SRE}
The stabilizer $\alpha$-Rényi entropy ($\alpha$-SRE) \cite{LeoneOlivieroHamma:2021} is a measure of non-stabilizerness,
defined in terms of (moments of) expectation values of Pauli strings as 
\begin{equation}
	M_{\alpha}(\ket{\psi})
	\coloneqq
	\frac{1}{1-\alpha} \log_{2}
	\bigg[
		\frac{1}{2^{N}} \sum_{P\in \{I,X,Y,Z\}^{\otimes N}} \big|\!\braket{\psi|P|\psi}\!\big|^{2\alpha}
	\bigg]
	\label{eq:SRE_definition}
\end{equation}
for an $N$-qubit state $\ket{\psi}$.
A naive computation requires a sum over $4^{N}$ Pauli strings in general,
and thus becomes computationally prohibitive as $N$ increases.
However, for integer~$\alpha$, the computation can potentially be simplified by the following formulation.
For an arbitrary $N$-qubit quantum state
\begin{equation}
	\ket{\psi}
	=
	\sum_{x\in \{0,1\}^{N}}c_{x} \ket{x},
\end{equation}
we introduce $2\alpha$ ``replicas'' and
(noting that $\big|\!\braket{\psi|P|\psi}\!\big|^{2\alpha} = \braket{\psi|P|\psi}^{2\alpha}$ since $P$ is Hermitian and $2\alpha$ is an even integer) write
\begin{align}
	\braket{\psi|P|\psi}^{2\alpha}
	& =
	\prod_{r=1}^{2\alpha}
	\sum_{x'_{(r)}\in \{0,1\}^{N}}
	\sum_{x_{(r)}\in \{0,1\}^{N}}
	c^{\ast}_{x'_{(r)}}
	c_{{x_{(r)}}}
	\braket{x'_{(r)}|P|x_{(r)}}
	\nonumber\\
	& =
	\sum_{\{x'_{(r)}\}_{r=1, \cdots, 2\alpha}}
	\sum_{\{x_{(r)}\}_{r=1, \cdots, 2\alpha}}
	\prod_{r=1}^{2\alpha}
	c^{\ast}_{x'_{(r)}}
	c_{x_{(r)}}
	\braket{x'_{(r)}|P|x_{(r)}}
	\nonumber\\
	& =
	\sum_{\{x'_{(r)}\}_{r=1, \cdots, 2\alpha}}
	\sum_{\{x_{(r)}\}_{r=1, \cdots, 2\alpha}}
	\prod_{r=1}^{2\alpha}
	c^{\ast}_{x'_{(r)}}
	c_{x_{(r)}}
	\prod_{i=1}^{N}
	\braket{x'_{(r),i}|P_i|x_{(r),i}}.
\end{align}
Then, one can take the sum over Pauli strings first as
\begin{equation}
	\sum_{P}
	\braket{\psi|P|\psi}^{2\alpha}
	=
	\sum_{\{x'_{(r)}\}_{r=1, \cdots, 2\alpha}}
	\sum_{\{x_{(r)}\}_{r=1, \cdots, 2\alpha}}
	\bigg(
		\prod_{r=1}^{2\alpha}
		c^{\ast}_{x'_{(r)}}
		c_{x_{(r)}}
	\bigg)
	\prod_{i=1}^{N}
	\sum_{P_i \in \{I,X,Y,Z\}}
	\prod_{r=1}^{2\alpha}
	\braket{x'_{(r),i}|P_i|x_{(r),i}}.
\end{equation}
Here,
%\begin{equation}
	$\prod_{r}
	\braket{x'_{(r),i}|P_i|x_{(r),i}}$
%\end{equation}
is non-zero only if either of the following holds:
\begin{itemize}
\item $P_i = I,Z$ and $x'_{(r),i} = x_{(r),i}$ for all $r$, or
\item $P_i = X,Y$ and $x'_{(r),i} \neq x_{(r),i}$ for all $r$.
\end{itemize}
As a result, for each $i$ we have
\begin{equation}
	\sum_{P_i \in \{I,X,Y,Z\}}
	\prod_{r=1}^{2\alpha}
	\braket{x'_{(r),i}|P_i|x_{(r),i}}
	=
	\begin{cases}
		1 + (-1)^{\#\{r\,|\,x_{(r),i} = 1\}} & \text{if }x'_{(r),i} = x_{(r),i}\text{ for all }r,\\
		1 + (-1)^{\#\{r\,|\,x_{(r),i} = 1\} + \alpha} & \text{if }x'_{(r),i} \neq x_{(r),i}\text{ for all }r,\\
		0 & \text{otherwise}.
	\end{cases}
\end{equation}

Let us now define the following set of pairs of $2\alpha$-tuples of configurations:
\begin{equation}
	S
	\coloneqq
	\left\{
		\Big(\{x_{(r)}\}, \{x'_{(r)}\}\Big)\,
	\middle|
	\begin{tabular}{l}
		for all $i$, either\\
		\phantom{or} $x'_{(r),i} = x_{(r),i}$ for all $r$ and $\#\{r\,|\,x_{(r),i} = 1\}$\,\phantom{$+\,\alpha$} is even\\
		or
		$x'_{(r),i} \neq x_{(r),i}$ for all $r$ and $\#\{r\,|\,x_{(r),i} = 1\}$\,$+\,\alpha$ is even
	\end{tabular}
	\right\}.
	\label{eq:set_S_definition}
\end{equation}
Then one can see that
\begin{equation}
	\prod_{i=1}^{N}
	\sum_{P_i \in \{I,X,Y,Z\}}
	\prod_{r=1}^{2\alpha}
	\braket{x'_{(r),i}|P_i|x_{(r),i}}
	=
	\begin{cases}
		2^{N} & \text{if }\Big(\{x_{(r)}\}, \{x'_{(r)}\}\Big) \in S,\\
		0 & \text{otherwise},
	\end{cases}
\end{equation}
and the stabilizer Rényi entropy simply reduces to
\begin{equation}
	M_{\alpha}\ket{\psi}
	=
	\frac{1}{1-\alpha}
	\log_{2}
	\sum_{(\{x_{(r)}\}, \{x'_{(r)}\}) \in S}
	\bigg(
		\prod_{r=1}^{2\alpha}
		c^{\ast}_{x'_{(r)}}
		c_{x_{(r)}}
	\bigg)
	\ .
	\label{eq:sre_reduced}
\end{equation}

%\newpage

%%%%%%%%%%%%%%%%%%%%%%%%%%%%%%%%%%%%%%%%%%%%%%%%%%%%%
\subsection{Dyck-Fredkin spin chain}\label{subsec:Dyck-Fredkin}
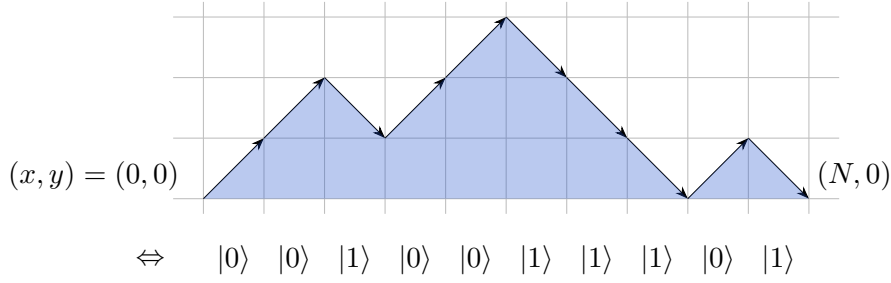
\begin{figure}[h]
	\centering
	
	\begin{tikzpicture}
		\node at (-1.05,0.3) {$(x,y) = (0,0)$};
		\node at (9.0,0.3) {$(N,0)$};
		\foreach \x in {0,...,3} {
			\draw[lightgray] (0, 0.8*\x) -- (8.8, 0.8*\x);
		}
		\foreach \x in {0,...,10} {
			\draw[lightgray] (0.4 + 0.8*\x,-0.2) -- (0.4 + 0.8*\x, 2.6);
		}
		\draw[-Stealth] (0.4,0.0) -- (1.2,0.8);
		\draw[-Stealth] (1.2,0.8) -- (2.0,1.6);
		\draw[-Stealth] (2.0,1.6) -- (2.8,0.8);
		\draw[-Stealth] (2.8,0.8) -- (3.6,1.6);
		\draw[-Stealth] (3.6,1.6) -- (4.4,2.4);
		\draw[-Stealth] (4.4,2.4) -- (5.2,1.6);
		\draw[-Stealth] (5.2,1.6) -- (6.0,0.8);
		\draw[-Stealth] (6.0,0.8) -- (6.8,0.0);
		\draw[-Stealth] (6.8,0.0) -- (7.6,0.8);
		\draw[-Stealth] (7.6,0.8) -- (8.4,0.0);
		\node at (-0.3, -0.8) {$\Leftrightarrow$};
		\node at (0.8, -0.8) {$\ket{0}$};
		\node at (1.6, -0.8) {$\ket{0}$};
		\node at (2.4, -0.8) {$\ket{1}$};
		\node at (3.2, -0.8) {$\ket{0}$};
		\node at (4.0, -0.8) {$\ket{0}$};
		\node at (4.8, -0.8) {$\ket{1}$};
		\node at (5.6, -0.8) {$\ket{1}$};
		\node at (6.4, -0.8) {$\ket{1}$};
		\node at (7.2, -0.8) {$\ket{0}$};
		\node at (8.0, -0.8) {$\ket{1}$};
		\fill[color=blue, opacity=0.3] (0.4,0.0) -- (2.0,1.6) -- (2.8,0.8) -- (4.4,2.4) -- (6.8,0) -- (7.6,0.8) -- (8.4,0) -- cycle;
	\end{tikzpicture}
	
	\caption{
		An example of a Dyck path for $N=10$
		and the corresponding spin/qubit configuration.
		The shaded region represents the area of the path, $A(p) = \sum_{i=0}^{N-1} \frac{1}{2}(y_i + y_{i+1}) = \sum_{i=1}^{N-1}y_i$.
	}
	
	\label{fig:Dyck_path}
\end{figure}

%%%%%%%%%%%%%%%%%%%%%%%%%%%%%%%%%%%%%%%%%%%%%%%%%%%%%
The Dyck-Fredkin model \cite{DellAnnaSalbergerBarbieroTrombettoniKorepin:2016, SalbergerKorepin:2016}
is a one-dimensional spin-$\frac{1}{2}$ chain
whose ground state is described in terms of combinatorial objects known as Dyck paths.
First, defining the set of lattice paths from $(0,0)$ to $(N,0)$ as
\begin{equation}
	\mathcal{P}_N
	\coloneqq
	\left\{
		(x_i,y_i)_{i=0,\cdots,N}
	\middle|
		\begin{array}{cccl}
			(x_0,y_0) & = & (0,0)\\
			(x_N,y_N) & = & (N,0)\\
			(x_{i+1},y_{i+1}) & = & (x_i+1, y_i \pm 1) & (0\leq i \leq N-1)
		\end{array}
	\right\},
\end{equation}
 a Dyck path is defined to be a path that stays on or above the $x$ axis, that is, an element of
\begin{equation}
	\mathcal{P}_N^{(\text{Dyck})}
	\coloneqq
	\left\{
		\vphantom{\bigg|}
		(x_i,y_i)_{i=0,\cdots,N} \in \mathcal{P}_N
	\,\middle|\,
		\begin{array}{cl}
			y_i \geq 0 &
			(0 \leq i \leq N)
		\end{array}
	\right\}.
\end{equation}
Then, by identifying each lattice path $p$ with a spin/qubit configuration $\ket{p}$ according to
\begin{equation}
	\left.
	\begin{array}{ccl}
		\nearrow\ : (x_{i+1},y_{i+1}) - (x_{i},y_{i}) = (+1,+1) & \leftrightarrow & \ket{\uparrow}/\ket{0}\\
		\searrow\ : (x_{i+1},y_{i+1}) - (x_{i},y_{i}) = (+1,-1)  & \leftrightarrow & \ket{\downarrow}/\ket{1}
	\end{array}
	\right\}
	\eqqcolon \ket{p_i}
\end{equation}
(see Fig.~\ref{fig:Dyck_path}), the ground state of the model is given by
\begin{equation}
	\ket{\psi_0^{\text{DF}}}
	=
	\frac{1}{\sqrt{|\mathcal{P}_N^{(\text{Dyck})}|}}
	\sum_{p\in \mathcal{P}_N^{(\text{Dyck})}}
	\underbrace{
		\bigotimes_i \ket{p_i}
	}_{
		=\ket{p}
	}.
\end{equation}
Note that the total number of length-$N$ Dyck paths is given by the Catalan number as
\begin{equation}
	|\mathcal{P}_N^{(\text{Dyck})}|
	=
	\begin{cases}
		\displaystyle
		C_n
		=
		\binom{2n}{n}
		-
		\binom{2n}{n-1}
		=
		\frac{1}{n+1}
		\binom{2n}{n}
%		\sim
%		\frac{4^n}{n^{3/2}\sqrt{\pi}}
		&
		N\text{ even }(N=2n),\\
		0
		&
		N\text{ odd}.\\
	\end{cases}
\end{equation}

The Dyck-Fredkin model also admits a deformation~\cite{SalbergerUdagawaZhangKatsuraKlichKorepin:2016}
whose ground state is given by a weighted superposition of Dyck-path states
\begin{equation}
	\ket{\psi_0^{\text{DF}}(t)}
	=
	\frac{1}{\sqrt{C(t)}}
	\sum_{p\in \mathcal{P}_N^{(\text{Dyck})}}
	t^{\frac{A(p)}{2}}\ket{p},
\end{equation}
where $C(t) = \sum_{p} t^{A(p)}$ is the normalization factor and $A(p)$ is the area between the path and the $x$~axis,
as shown in Fig.~\ref{fig:Dyck_path}.
The original undeformed model corresponds to $t=1$. 
For $t<1$ (resp. $t>1$), Dyck paths with smaller (resp. larger) areas have larger weights in the ground state.
In the limits $t\to 0$ and $t\to\infty$, the ground state approaches
the minimum-area path state $\ket{01\cdots 01}$ and
the maximum-area path state  $\ket{0\cdots01\cdots 1}$, respectively,
both of which are stabilizer states.

Since our analysis below relies only on the form of the ground state,
we omit further details, including the explicit form of the Hamiltonian.
The interested reader is referred to the original papers cited above.

%\newpage

%%%%%%%%%%%%%%%%%%%%%%%%%%%%%%%%%%%%%%%%%%%%%
\section{Formulation and analysis}\label{sec:formulation}
%%%%%%%%%%%%%%%%%%%%%%%%%%%%%%%%%%%%%%%%%%%%%
We now apply the replica formulation of the SRE to the ground state of the (deformed) Dyck-Fredkin spin chain.
Simply substituting
\begin{equation}
	c_x = \begin{cases}
		\dfrac{1}{\sqrt{C(t)}}\,t^{\frac{A(x)}{2}} & \text{if }x\in P_{N}^{(\text{Dyck})}\\[12pt]
		\qquad 0 & \text{otherwise}
	\end{cases}
\end{equation}
into Eq.~\eqref{eq:sre_reduced}, we obtain
\begin{equation}
	M_{\alpha}(\ket{\psi_0^{\text{DF}}(t)})
	=
	\frac{1}{1-\alpha}\log_{2}\Bigg[
		\frac{1}{[C(t)]^{2\alpha}}
		\sum_{
			\substack{
				(\{x_{(r)}\}, \{x'_{(r)}\}) \in S\\
				\forall r,\  
				x_{(r)}, x'_{(r)}\in \mathcal{P}_N^{(\text{Dyck})}
			}
		}
		t^{\frac{1}{2}\sum_{r=1}^{2\alpha}[A(x_{(r)}) + A(x'_{(r)})]}
	\Bigg].
	\label{eq:Dyck_Fredkin_GS_SRE}
\end{equation}
For even $\alpha$, the constraints defining the set $S$ in Eq.~\eqref{eq:set_S_definition} simplify,
and the quantity inside the brackets can be computed efficiently using dynamic programming,
without explicitly enumerating all $\mathcal{O}(2^{N\cdot 4\alpha})$ configurations of the replicated Dyck paths.
This is because it is sufficient to keep track only of the ``heights'' $y_{i}$ of the replicated paths ($\mathcal{O}(N^{4\alpha})$ states) at each step~$i$.
This allows us to perform exact computations for substantially larger system sizes than is feasible with a direct evaluation of Eq.~\eqref{eq:SRE_definition};
we verified our implementation against the latter at $\alpha=2$ for $N \leq 8$ and several values of $t$, finding complete agreement to within numerical precision.

From Eq.~\eqref{eq:Dyck_Fredkin_GS_SRE}, we can also infer the asymptotic behavior of the SRE with respect to $t$.
To this end, first note that
\begin{equation}
	C(t) = 1 \cdot t^{A_{\text{min.}}} + \left(\frac{N}{2}-1\right)\cdot t^{A_{\text{min.}}+2} + \cdots + 1 \cdot t^{A_{\text{max.}}-2} + 1 \cdot t^{A_{\text{max.}}}
\end{equation}
where $A_{\text{min.}} = \frac{N}{2}$ and $A_{\text{max.}} = \frac{N^2}{4}$.
For $t\to0$, since the constraint in Eq.~\eqref{eq:set_S_definition} does not allow a configuration
in which only a single replica pair differs from the minimum-area Dyck path, 
\begin{align}
	M_{\alpha}(\ket{\psi_0^{\text{DF}}(t)})
	& =
	\frac{1}{1-\alpha}\log_{2} \bigg[
		\underbrace{
			\frac{1}{[C(t)]^{2\alpha}}
			\cdot t^{2\alpha \cdot A_{\text{min.}}}
		}_{
			= [1 + (\frac{N}{2}-1)\cdot t^2 + \mathcal{O}(t^4)]^{-2\alpha}
		} \Big(1 + \mathcal{O}(t^3)\Big)
	\bigg]
	\notag\\[4pt]
	& =
	-\frac{1}{\log_{e}2} \cdot \frac{2\alpha}{1-\alpha}\left(\frac{N}{2}-1\right) t^2 + \mathcal{O}(t^3).
\end{align}
For $t\to\infty$, a configuration
in which only a single replica pair differs from the maximum-area Dyck path is again not allowed, leading to 
\begin{align}
	M_{\alpha}(\ket{\psi_0^{\text{DF}}(t)})
	& =
	\frac{1}{1-\alpha}\log_{2}\bigg[
		\underbrace{
			\frac{1}{[C(t)]^{2\alpha}}
			\cdot t^{2\alpha \cdot A_{\text{max.}}}
		}_{
			= [1 + 1\cdot t^{-2} + \mathcal{O}(t^{-4})]^{-2\alpha}
		} \Big(1 + \mathcal{O}(t^{-3})\Big)
	\bigg]
	\notag\\[4pt]
	& =
	-\frac{1}{\log_{e}2} \cdot \frac{2\alpha}{1-\alpha} t^{-2} + \mathcal{O}(t^{-3}).
\end{align}
The contrasting $N$-dependence in the two limits thus derives from
the difference in the number of sub-extremal-area Dyck paths.

%\newpage

%%%%%%%%%%%%%%%%%%%%%%%%%%%%%%%%%%%%%%%%%%%%%
\section{Results}\label{sec:results}
%%%%%%%%%%%%%%%%%%%%%%%%%%%%%%%%%%%%%%%%%%%%%
Here, we present our results on how the stabilizer Rényi entropy scales with system size
across the different parameter regions of the (deformed) Dyck-Fredkin spin chain.

\begin{figure}[h]
	\centering
	
	\captionsetup[subfigure]{skip=0pt}
	
	\begin{subfigure}[h]{\textwidth}
		\centering
		\includegraphics[width=0.39\linewidth]{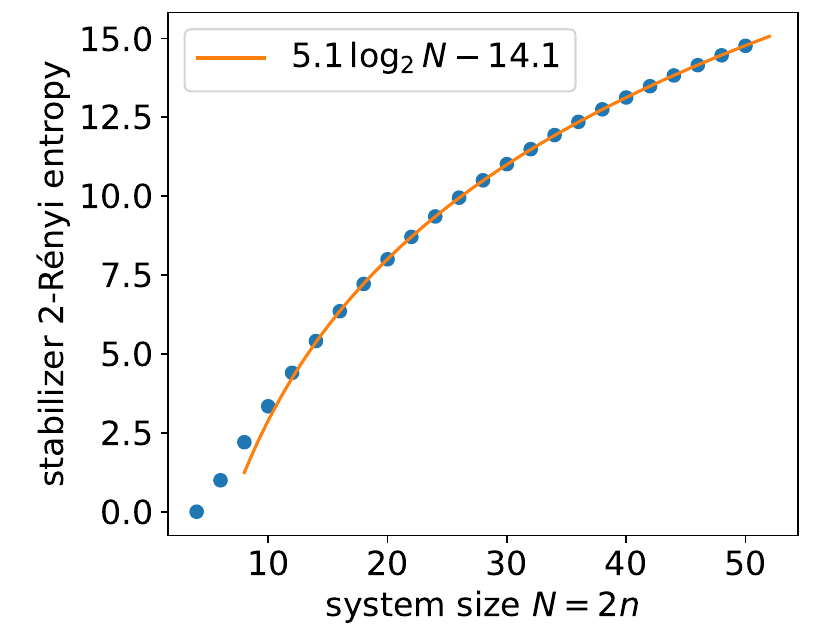}
		\includegraphics[width=0.39\linewidth]{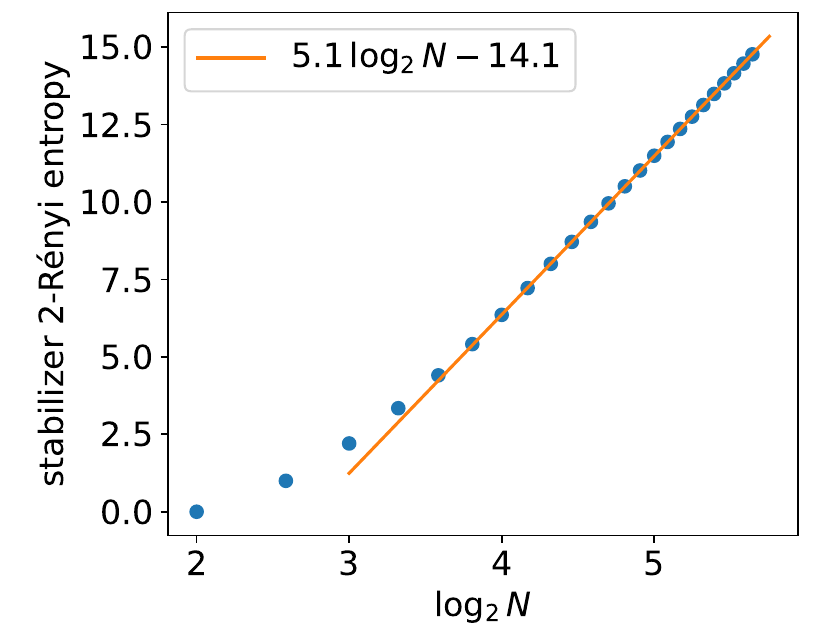}
		\caption{$\alpha=2$\vspace{14pt}}
	\end{subfigure}
	
	\begin{subfigure}[h]{\textwidth}
		\centering
		\includegraphics[width=0.39\linewidth]{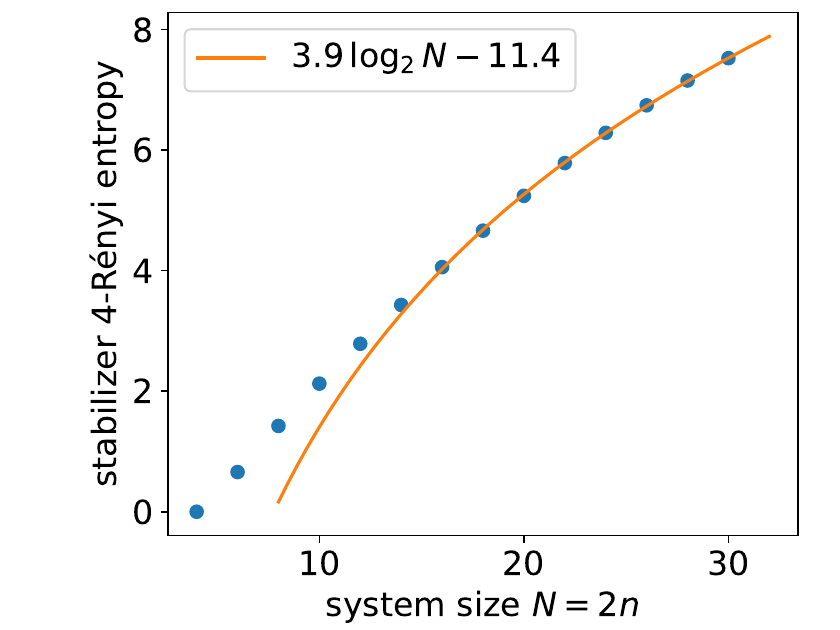}\,
		\includegraphics[width=0.39\linewidth]{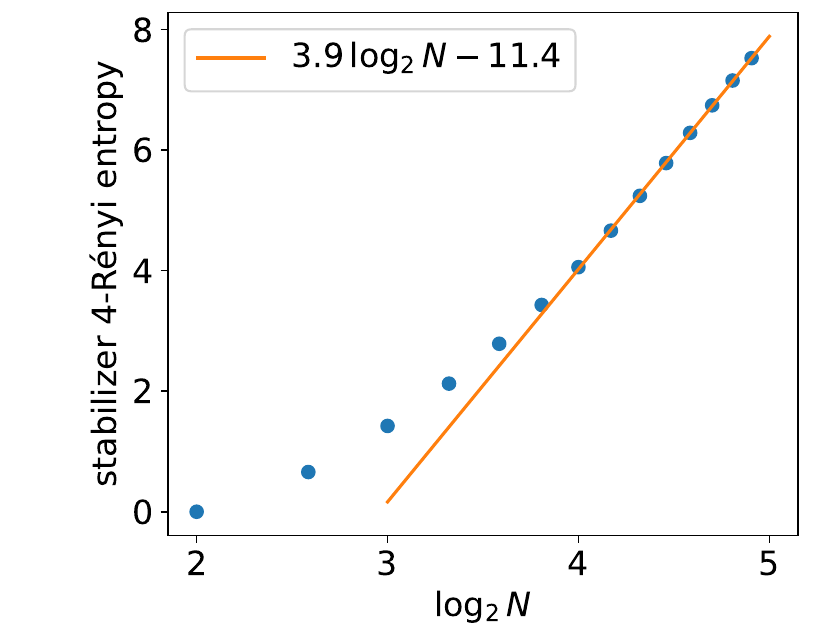}
		\caption{$\alpha=4$}
	\end{subfigure}
	
	\caption{
		Stabilizer $\alpha$-Rényi entropy $M_{\alpha}(\ket{\psi_0^{\text{DF}}(1)})$
		of the ground state of the (undeformed) Dyck-Fredkin spin chain
		as a function of system size $N$. 
	}
	
	\label{fig:undeformed_SRE}
\end{figure}

Let us first consider the undeformed case.
Figure~\ref{fig:undeformed_SRE} shows the system-size dependence of the stabilizer $\alpha$-Rényi entropy for $\alpha=2$ and $4$.
For both values of $\alpha$, the SRE exhibits logarithmic scaling with system size,
suggesting that this behavior is not an artifact of a particular value of $\alpha$.

%\newpage

\begin{figure}[h]
	\centering
	
	\begin{subfigure}[h]{0.32\textwidth}
		\includegraphics[width=\linewidth]{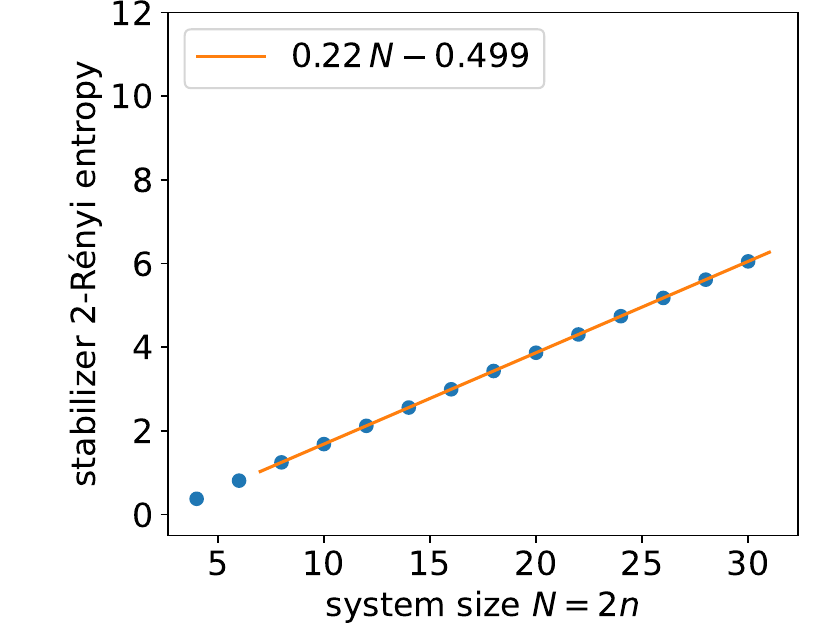}
		\caption{$t=0.5$}
	\end{subfigure}
	\begin{subfigure}[h]{0.32\textwidth}
		\includegraphics[width=\linewidth]{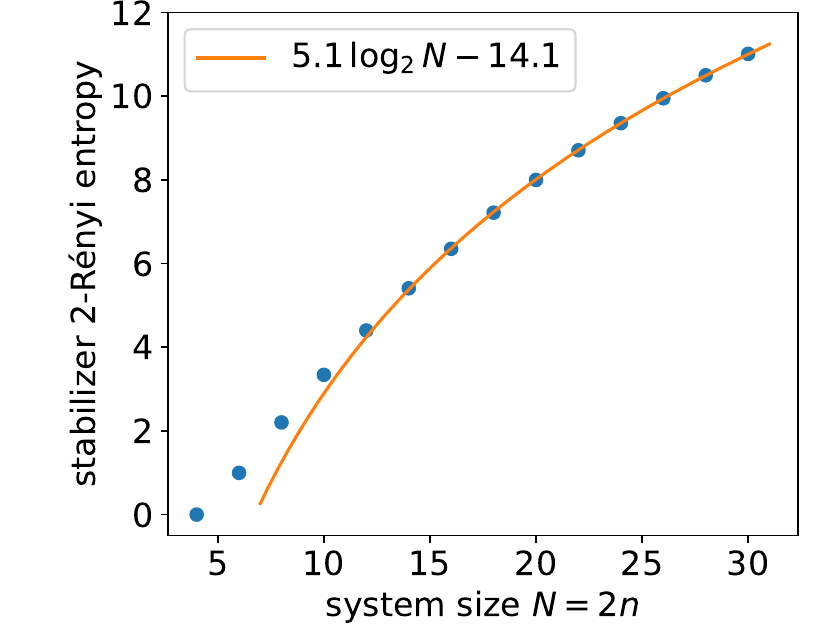}
		\caption{$t=1$}
	\end{subfigure}
	\begin{subfigure}[h]{0.32\textwidth}
		\includegraphics[width=\linewidth]{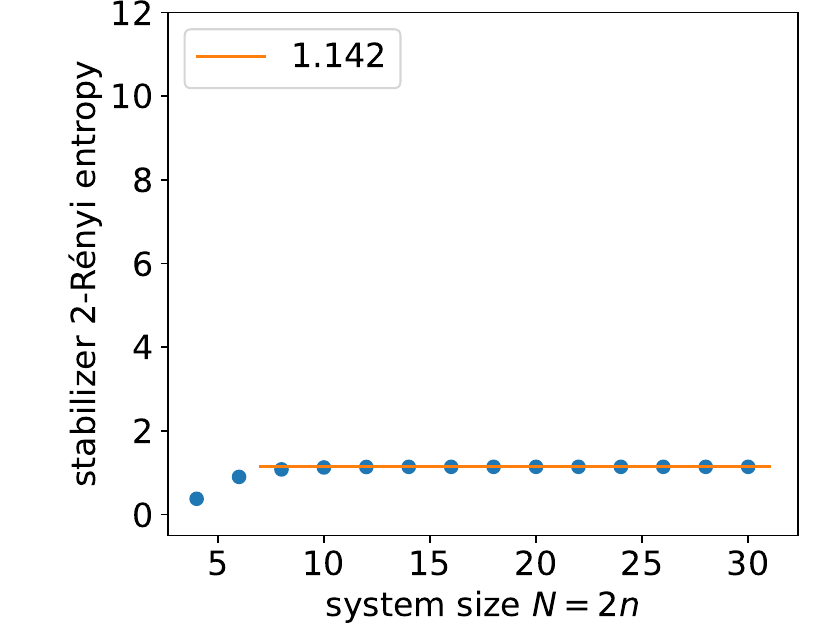}
		\caption{$t=2$}
	\end{subfigure}
	
	\caption{
		Stabilizer $2$-Rényi entropy $M_{2}(\ket{\psi_0^{\text{DF}}(t)})$ of
		the ground state of the deformed Dyck-Fredkin spin chain
		as a function of system size $N$. 
	}
	
	\label{fig:deformed_SRE}
\end{figure}

We next examine how this scaling behavior changes under deformation.
Figure~\ref{fig:deformed_SRE} shows the system-size dependence of the 2-SRE 
at representative values of the deformation parameter.
The results exhibit three qualitatively distinct scaling behaviors:
linear $(t<1)$, logarithmic $(t=1)$, and constant~$(t>1)$.
The behavior away from $t=1$ is indeed consistent with the asymptotic analysis in Sec.~\ref{sec:formulation}.
The SRE vanishes in both limits $t\to 0$ and $t\to\infty$,
again consistent with the fact that the corresponding ground states are stabilizer states.
Our numerical results also agree with the expected tendencies toward this vanishing:
a decreasing slope of the linear scaling as $t\to 0$ and
a decreasing magnitude as~$t\to \infty$.

%%%%%%%%%%%%%%%%%%%%%%%%%%%%%%%%%%%%%%%%%%%%%
\section{Discussion}\label{sec:discussion}
%%%%%%%%%%%%%%%%%%%%%%%%%%%%%%%%%%%%%%%%%%%%%
The results in the previous section are intriguing when compared with
the known spectral and entanglement properties of the model,
as summarized in Table~\ref{table:comparison}.

\begin{table}[b]
	\centering
	
	\caption{Known (and numerically suggested) scaling behaviors across the three parameter regions.}
	
	\begin{tabular}{ccccc}
		& $t<1$ & $t=1$ & $t>1$\\
		\hline
		spectral gap & $\Theta(1)$ & $\propto N^{-z}$ $(z\approx 3)$ & $\mathcal{O}(t^{-N})$\\
		entanglement entropy & $\Theta(1)$ & $\Theta(\log N)$& $\Theta(1)$\\
		SRE & $\Theta(N)$ & $\Theta(\log N)$& $\Theta(1)$
	\end{tabular}
	
	\label{table:comparison}
\end{table}

At the undeformed point $t=1$, it is known that
the spectral gap $\Delta E$ closes polynomially as $\Delta E\propto N^{-z}$
with an unusual exponent $z \geq 2$~\cite{Movassagh:2016},
and numerical studies further indicate $z \approx 3$ \cite{DellAnnaSalbergerBarbieroTrombettoniKorepin:2016, ChenFradkinWitczakKrempa:2017,AdhikariBeach:2020MC}.
While the asymptotic scaling of the entanglement entropy~\cite{SalbergerUdagawaZhangKatsuraKlichKorepin:2016} is the same as that at conventional critical points,
the anomalous logarithmic scaling of the SRE occurs precisely at this point.
Together with the observation that the W~state \cite{OdavićHaugTorreHammaFranchiniGiampaolo:2022} and the half-filled Dicke state \cite{Lee:half_filled_Dicke},
both lying in the degenerate ground-state manifold of the spin-$\frac{1}{2}$ ferromagnetic Heisenberg XXX chain with $z=2$,
also exhibit logarithmic SRE scaling,
it is tempting to speculate that there is a connection between non-stabilizerness and unconventional criticality.
Furthermore, along with the recent work \cite{HallamSmithPapić:2026},
it raises the broader question of how different measures of non-stabilizerness resolve critical properties of quantum many-body systems.

Under deformation, it is notable that
the entanglement entropy exhibits the same (constant) scaling for both $t<1$ and $t>1$ \cite{MooreBeach:2025},
despite the qualitatively different spectral-gap behavior \cite{UdagawaKatsura:2017,AdhikariBeach:2020DMRG},
whereas the SRE at least apparently distinguishes the two regimes.
Whether this is a mere coincidence remains unclear at the moment,
but it lends further support to the view that non-stabilizerness may encode information about the system
that is not captured by entanglement~\cite{PassarelliFazioLucignano:2024},
motivating further investigation.
We leave extensions to the colored Dyck-Fredkin and (colored) Motzkin chains for future work.

Lastly, note that
lattice path enumeration is closely related to random walks,
and the replica formulation of the SRE can indeed be interpreted as a random walk on $(\mathbb{Z}_{\geq 0})^{4\alpha}$.
Accordingly, asymptotic results for random walks in cones (see, e.g., \cite{DenisovWachtel:2015})
may provide a route toward a more detailed understanding of the logarithmic scaling observed at the undeformed point.

\newpage

\thispagestyle{empty}

\section*{Acknowledgments}
The author is deeply grateful to
Leonard Logari\'c
for drawing his attention to the Dyck-Fredkin spin chain and sharing helpful information.
The author would also like to thank Nobuyuki Yoshioka, as well as ICEPP,
for offering the opportunity and creating an environment where this work was~possible.
The author is also indebted to Nobuyuki Yoshioka for valuable comments on an earlier draft.
The author is supported by JST [Moonshot R\&D Program] Grant No. [JPMJMS256J].

\bibliographystyle{ytamsalpha}
\bibliography{bib}
\thispagestyle{empty}

\end{document}